\documentclass[dvipsnames]{fairmeta}
\usepackage{microtype}

\usepackage{booktabs} % for professional tables
\usepackage{algorithmic}
\usepackage[vlined,ruled,commentsnumbered,linesnumbered]{algorithm2e}
\usepackage{tabularx}
\usepackage{multirow}
\usepackage{marvosym}
\usepackage{pifont}
\usepackage{float}
\usepackage{listings}
\usepackage{wasysym}
\usepackage{subcaption}
\usepackage[justification=centering]{caption}
\usepackage{xcolor}

\usepackage[T1]{fontenc}
\usepackage{inconsolata}
\definecolor{commentcolor}{RGB}{100,100,100}
\definecolor{keywordcolor}{RGB}{180,20,20}
\definecolor{stringcolor}{RGB}{60,110,60}
\definecolor{shadegray}{RGB}{245,245,245}

\lstdefinestyle{pythonstyle}{
    language=Python,
    basicstyle=\fontfamily{pcr}\fontsize{6.5}{9.6}\fontseries{mb}\fontencoding{T1}\fontseries{mb}\selectfont,  % Consolas at 8pt with 1.2x line spacing
    backgroundcolor=\color{shadegray},
    numbers=left,
    numberstyle=\tiny\color{gray},
    numbersep=5pt,
    frame=none,
    breaklines=true,
    keywordstyle=\color{keywordcolor},
    commentstyle=\color{commentcolor}\itshape,
    stringstyle=\color{stringcolor},
    showstringspaces=false,
    tabsize=4,
    morekeywords={self,def,class,return,None,raise,from,import,as,for,while,if,else,elif,try,except,True,False},
    literate={*}{{\char42}}1
            {-}{{\char45}}1
            {+}{{\char43}}1
            {=}{{\char61}}1
            {>}{{\char62}}1
            {<}{{\char60}}1,
    emphstyle=\color{keywordcolor},
    emph={ReplicateComputation,Replicate,Partial,CheckpointPolicy},
    columns=flexible,
    keepspaces=true,
}
\usepackage{hyperref}
\usepackage{subfloat}
\newcommand{\cmark}{\ding{51}}%
\newcommand{\sys}{SimpleFSDP}
\newcommand{\compiler}{TorchInductor}

\title{\sys{}: Simpler Fully Sharded Data Parallel with torch.compile}

\author[1,*,\dagger]{Ruisi Zhang}
\author[2,*]{Tianyu Liu}
\author[2]{Will Feng}
\author[2]{Andrew Gu}
\author[3,\dagger]{Sanket Purandare}
\author[2]{Wanchao Liang}
\author[2]{Francisco Massa}

\affiliation[1]{UC San Diego}
\affiliation[2]{Meta}
\affiliation[3]{Harvard University}

\contribution[*]{Equal contribution}
\contribution[\dagger]{Work done at Meta}

\abstract{Distributed training of large models consumes enormous computation resources and requires substantial engineering efforts to compose various training techniques.
This paper presents \sys{}, a PyTorch-native compiler-based Fully Sharded Data Parallel (FSDP) framework, which has a simple implementation for maintenance and composability, allows full computation-communication graph tracing, and brings performance enhancement via compiler backend optimizations. 

\sys{}'s novelty lies in its unique \texttt{torch.compile}-friendly implementation of collective communications using existing PyTorch primitives, namely parametrizations, selective activation checkpointing, and DTensor. It also features the first-of-its-kind intermediate
representation (IR) nodes bucketing and reordering in the \compiler{} backend for effective computation-communication overlapping. 
As a result, users can employ the aforementioned optimizations to automatically or manually wrap model components for minimal communication exposure. Extensive evaluations of \sys{} on Llama 3 models (including the ultra-large 405B) using TorchTitan demonstrate up to 28.54\% memory reduction and 68.67\% throughput improvement compared to the most widely adopted FSDP2 eager framework, when composed with other distributed training techniques.}

\date{\today}
\correspondence{Ruisi Zhang at \email{ruz032@ucsd.edu}, Tianyu Liu at \email{lty@meta.com}}

\begin{document}

\maketitle

\section{Introduction}
% what is the problem
Distributed training the ever-growing large models necessitates huge computation resources~\cite{rae2021scaling,zhang2022opt,chowdhery2023palm,dubey2024llama} and engineering efforts~\cite{shoeybi2019megatron,rasley2020deepspeed,Liang2024TorchTitanOP}, both of which pose significant challenges as the model size scales.  
For example, training the Llama 3.1 405B~\cite{dubey2024llama} model takes 30.84 million H100 GPU hours, and PaLM-540B~\cite{chowdhery2023palm} model takes 9.4 million TPUv4 hours. 
During training, various parallelisms~\cite{huang2019gpipe,shoeybi2019megatron,zhao2023pytorch}, memory optimizations~\cite{chen2016training,korthikanti2023reducing}, and communication optimizations~\cite{micikevicius2017mixed,zhao2023pytorch,choudhury2024mast} are employed to improve computation throughputs and minimize communication exposure. 
%Various parallelism and communication optimizations 

Fully Sharded Data Parallel (FSDP)~\cite{zhao2023pytorch}, motivated by the DeepSpeed ZeRO~\cite{rajbhandari2020zero}, is one of the most fundamental techniques for distributed large model training. It significantly saves memory by sharding model parameters, gradients, and optimizer states across multiple devices and only gathers them when needed. As such, it is widely adopted to train large generative models~\cite{le2023bloom,dubey2024llama} and has been deployed in open-source libraries, like NeMo~\cite{kuchaiev2019nemo}, DeepSpeed~\cite{rasley2020deepspeed}, and TorchTitan~\cite{Liang2024TorchTitanOP}.

%in open-source libraries, like Megatron~\cite{shoeybi2019megatron} and DeepSpeed~\cite{rasley2020deepspeed} and used for many generative models training like Llama~\cite{dubey2024llama}. FSDP shards the  and only gathers the parameters needed for computation and discards them immediately after using them to save a significant amount of memory.
%It is usually composed with other efficient training techniques to train ultra-large models.

FSDP is primarily developed in the PyTorch \emph{eager} (i.e., non-compile) mode, where model operators are executed immediately after definition. It preserves debuggability and enables certain mechanisms like pre-fetching via backward hooks~\cite{hook-rfc}, which are hard to trace in the compile mode~\cite{ansel2024pytorch}. However, the eager mode impairs the training performance, as the model cannot be compiled as a whole graph, thereby losing opportunities for hardware-specific computation optimizations and efficient memory management.

% how prior work solve this
Prior works bringing machine learning compilers into distributed training mainly go from two directions: (1) JAX-based~\cite{jax2018github,xu2021gspmd} that uses XLA~\cite{sabne2020xla} as the compile backend and shard tensors via user annotations; (2) PyTorch-based~\cite{Liang2024TorchTitanOP}, which leverages \texttt{torch.compile}~\cite{ansel2024pytorch} to trace per-device compute submodules and insert inter-module communications.
JAX adopts functional programming and imposes certain constraints to ensure compatibility with the XLA backend. This greatly hinders the programmability in distributed training, which stacks many emerging techniques and requires agile development.

PyTorch-based approach~\cite{Liang2024TorchTitanOP}, on the other hand, only compiles the model's computation modules, as the FSDP eager-mode implementations like prefetching are hard to be traced by \texttt{torch.compile}.
Hence, it loses the opportunity to compile a full model graph for communication/computation co-optimization and introduces additional codebase complexity by requiring the manual insertion of inter-module communications.

%introduced additional complexity to insert inter-module communications into the codebase. 

% requires a lot of engineering effort to trace collective communications in FSDP. Certain mechanisms, like pre-fetchings, require backward hoods, which are hard to be traced by \texttt{torch.compile}.
% As a result, users have to manually wrap the modules they want to compile and insert collective communications in between. This greatly increased the complexity of the codebase and loses potential opportunities for co-optimizing the communication and computation overlapping to further enhance the training performance.

This paper presents \sys{}, a PyTorch-native compiler-based FSDP framework. It features 
(1) \textbf{Simplicity}: users do not need to alter the eager-mode distributed training codebase while experiencing the performance enhancement from full-model compilation; (2) \textbf{Composability}: \sys{} can be seamlessly integrated with emerging distributed training techniques with minimal engineering effort; (3) \textbf{Performance}: training throughputs and memory gains from full-graph tracing and compiler optimizations;  and (4) \textbf{Debuggability}: \sys{} exhibits usability in PyTorch eager mode, where users have the flexibility to debug and agile develop the codebase.

\sys{} achieves the FSDP semantics by utilizing a few existing PyTorch primitives. 
First, representing the sharded per-parameter as DTensors~\cite{dtensor-rfc}, \sys{} achieves the ``all-gather before usage'' behavior by applying collective communications (via the DTensor \verb|redistribute| API) as tensor parametrization. Note that the backward gradient reduce-scatter is automatically achieved as parametrization and DTensor \verb|redistribute| are differentiable. 
Second, given that in parametrization~\cite{parameterize-rfc}, parameter all-gathers are treated as activation computations, \sys{} achieves the additional memory optimization of ``release after forward usage, all-gather again before backward usage'' by wrapping the parametrization module using activation checkpointing~\cite{checkpoint-rfc}. Since parametrization, selective activation checkpointing, and DTensor APIs are all natively supported by \texttt{torch.compile}, \sys{} obtains a full graph of communication and computation operations.

%Identifying that communication operators in FSDP (ZeRO-3) have similar behaviors with activation checkpointing -- tensors are computed/gathered in the forward pass; they are re-computed/re-gathered in the backward pass. \sys{} parametrizes~\cite{parameterize-rfc} the all-gather to checkpoint~\cite{checkpoint-rfc} the parameters abstracted by DTensor~\cite{dtensor-rfc} in the model, which enables the parameters to be re-gathered in the backward pass. The parametrization, selective activation checkpointing, and DTensor APIs are PyTorch-native, enabling them to be traced by \texttt{torch.compile}. As a result, \sys{} obtains a full graph of communication and computation operations. 

\sys{} introduces two optimization components in \texttt{torch.compile}'s backend \compiler{}, namely bucketing and reordering, to enhance the per-parameter sharding performance. The bucketing merges the communication operations\footnote{The operators are lowered to IR nodes in \compiler{}. We use the two terms interchangeably throughout the paper.} in \compiler{} to reduce the frequency of issuing base communication. The reordering pre-fetches the parameters used for computation in later stages to overlap with the current stage's computation for minimized communication exposure. 
Building on top of the optimizations, \sys{} provides two interfaces to users to wrap the model, enabling both customization and automation. The first manual-wrapping enables users to customize the communications to bucket among modules and reorders the bucketed communication operations to reduce exposure. The auto-wrapping employs a greedy algorithm to bucket the communication operations as long as they can be overlapped by the computation operations and do not exceed memory limits. 

\sys's PyTorch-native implementation enables it to be seamlessly composed with other distributed training techniques. We demonstrate its composability with Tensor 
 Parallel and Pipeline Parallel, meta initialization, mixed precision training, and activation checkpointing with only a few lines of code. Such composability is achieved while tracing the model's full computation-communication graph and tested on scales up to the ultra-large 405 billion parameter Llama 3.1 model~\cite{dubey2024llama}.

% Tianyu: I still think the following statement is technically not clear. I suggest we remove it for now, before further discussion. 
% In addition, by formulating FSDP as a composition of two commonly-used primitives, we can reduce the search space for auto-parallelism strategies, as FSDP doesn't need to be separately taken into account.
% \textcolor{blue}{@Francisco, do you want to add the auto-parallelism and simplify fsdp operation part somewhere in the intro?}

In summary, our contributions are as follows:
\begin{itemize}

\item    We introduce \sys{}, a PyTorch-native compiler-based FSDP framework featuring simplicity, composability, performance enhancement, and debuggability. % , by decomposing FSDP into parameter-sharding and activation checkpointing of parameter-gathering.

\item  We devise \sys{} highlighting (1) a unique collective communication implementation of FSDP via PyTorch primitives (parametrizations, selective activation checkpointing, and DTensor API), enabling full-graph tracing in model training; (2) the first-of-its-kind IR nodes bucketing and reordering in \compiler{} with flexible user interfaces (manual-wrapping and auto-wrapping) to customize and automate computation-communication overlapping. 

\item  We perform extensive evaluations of \sys{} on Llama 3 models (up to the ultra-large 405B) using TorchTitan~\cite{Liang2024TorchTitanOP}, demonstrating its 
    (1) \textbf{Performance}: up to 28.54\% peak memory reduction and 68.67\% higher throughput improvement, compared to the most widely adopted FSDP2 eager framework~\cite{fsdp-rfc}; (2) \textbf{Scalability and Composability}: full-graph tracing when composed with other distributed training techniques while maintaining up to 6.06\% throughput gains and 8.37\% memory reduction, compared to the existing best-performing sub-module compilation; (3) \textbf{Debuggability}: maintaining comparable memory and throughput in the eager mode.
\end{itemize}

%As such, \sys{} keeps the \textbf{Simplicity} by handling the optimizations in \compiler{} instead of the distributed training codebase and \textbf{Composability} to compose with other efficient-training techniques with PyTorch-native implementation. 

\section{Background and Challenges}
This section first introduces techniques and related work for accelerating large models' distributed training. We then present several challenges that state-of-the-art frameworks face when supporting these techniques.

\subsection{Distributed Training Large Models}\label{subsec:dist_training}
Training large models in a distributed manner reduces the memory requirements per device and accelerates the computation throughputs. 

\textit{Fully Sharded Data Parallel} (FSDP)~\cite{zhao2023pytorch} is one of the most fundamental forms of data parallelism in distributed training. It shards model parameters, gradients, and optimizer states across multiple devices. During training, it gathers the needed parameters for computation and discards them immediately to save memory. 
A typical FSDP training consists of 

%\begin{itemize}
$\bullet$ \textbf{Model Initialization \& Parameter Sharding}: The model is wrapped into FSDP units and partitioned per the number of devices for parameter sharding. Each device only holds one of the partitions. 
 
$\bullet$ \textbf{Forward Pass}: Each FSDP unit gathers the parameters from other devices and performs the computation. The parameters are discarded immediately after the computation to save memory. 
 
$\bullet$  \textbf{Backward Pass}: Similar to the forward pass, each FSDP unit re-gathers the parameters and computes the gradient. The gradients are averaged and sharded across devices.  
%\end{itemize}

\textit{Tensor Parallel}~\cite{shoeybi2019megatron,narayanan2021efficient} partitions and shards tensors of an individual layer across multiple devices.  Each device computes the sharded part of the layer simultaneously and concatenates them together for the outputs. 
\textit{Pipeline Parallel}~\cite{huang2019gpipe,narayanan2019pipedream,li2021terapipe} partitions a model that cannot fit in a single device's memory into multiple stages. Each device concurrently processes a stage over multiple batches of data. 

\textit{Meta initialization}~\cite{zhao2023pytorch} initializes the model parameters on a meta device~\cite{meta-device-rfc} (an abstract device that denotes a tensor and records only metadata) rather than the actual CPU/GPU device. It reduces the time and memory required for initialization.  
\textit{Mixed precision training}~\cite{micikevicius2017mixed} reduces memory usage by training the model using 16-bit floating-point numbers. In the gradient updates, parameters are cast back to 32-bit floating-point numbers for training stability. 
\textit{Activation checkpointing}~\cite{chen2016training, korthikanti2023reducing} reduces the memory consumption by selectively storing activations at certain layers in the forward pass and recomputing the rest during the backward pass. It significantly reduces the peak memory and allows model training on memory-constraint devices.

\subsection{Related Work}

Machine learning compilers~\cite{chen2018tvm,sabne2020xla,ansel2024pytorch} accelerates model training by optimizing the computation graph execution on different target hardware devices and by performing careful memory management. In \texttt{torch.compile}~\cite{ansel2024pytorch}, the frontend TorchDynamo captures the FX graph from the user code by just-in-time (JIT) compiling Python bytecode; the default backend \compiler{} takes the FX graph operations as input and lowers the graph to a set of intermediate representation (IR) nodes.  The IR nodes are fused and generate corresponding  OpenAI Triton code~\cite{tillet2019triton} to write GPU kernels for more efficient execution.

Distributed training frameworks like Megatron-LM~\cite{shoeybi2019megatron} and DeepSpeed~\cite{rasley2020deepspeed} support various parallelism strategies (data, tensor, and pipeline) and memory-saving techniques (activation checkpointing and mixed precision training) to train large transformer language models at scale. They are primarily developed in the PyTorch eager mode for its debuggability and easy-to-use interface.
\sys{} is a PyTorch-native FSDP implementation, providing a simple plug-and-play interface, compiler-based optimizations, and composability with other parallelisms. Therefore, \sys{} is orthogonal to them and can be potentially integrated into those frameworks as enhancements.

Existing auto-parallelism works~\cite{zheng2022alpa,lin2024nnscaler,chen2024slapo} attempted to cover Data Parallel, but without optimizing computation-communication overlapping, thus potentially yielding sub-optimal solutions. We hope \sys{}'s systematic exploration of automating communication optimizations, as well as the infrastructure innovation in compiler backend, would benefit future auto-parallelism works.

\subsection{Challenges}
\label{subsec:challenge}

Applying machine-learning compilers to distributed training exhibits a few challenges, as outlined below.

\textbf{Complexity} Distributed training combines various parallelisms and memory-saving techniques to fit larger models and increase throughputs, making the codebase inherently complicated, especially when the techniques strives to improve eager-mode performance. This makes the integration and tracing of machine learning compilers difficult. 
% For example, in TorchTitan, users have to manually define shared components and compile the computations within them, as supporting collective communication tracing is still ongoing in \texttt{torch.compile}.
%; (ii) migrate their code to new machine learning frameworks like JAX. Both options introduce additional complexity to the codebase and create challenges for developers.

\textbf{Debuggability} While machine learning compilers offer performance enhancements, the debuggability from the eager mode remains crucial. It allows users to experiment with different building blocks for agile development. However, these debugging practices may violate the compilation rules, making the code untraceable~\cite{jax2018github,sabne2020xla}. As a result, a framework that preserves both debuggability under eager mode and performance enhancement from compile mode is essential.

\textbf{Composability} %Tracing the collective communication directly from the eager mode FSDP code requires huge engineering efforts. 
Existing parallelism implementations (e.g. DDP and FSDP in PyTorch) use backward hooks to perform efficient collective communications, making it difficult for \verb|torch.compile| to trace. Although attempts have been made to enable full-graph compilation in those scenarios, integrating them with emerging distributed training techniques are still challenging.

\section{\sys{} Design}
Identifying the challenges, we introduce \sys{}, which maintains distributed training's \textbf{simplicity} and \textbf{debuggability}. Then, we incorporate several compiler-only optimization components to enhance \sys{}'s \textbf{performance}.  We demonstrate \sys's \textbf{composability} with other distributed training techniques in Section~\ref{sec:composability}. %Additionally, and its \textbf{extensibility} to downstream applications, such as auto-parallelism.

\subsection{Overview}
\label{subsec:overview}

In this section, we introduce how \sys{} realizes the FSDP semantics using existing PyTorch primitives, namely \emph{parametrization}~\cite{parameterize-rfc} and \emph{selective activation checkpointing}~\cite{checkpoint-rfc}, together with the DTensor abstractions~\cite{dtensor-rfc}. All these techniques are now natively supported by PyTorch \texttt{torch.compile}.

\begin{figure}[!ht]
\begin{center}
    \centering
    \includegraphics[width=0.6\columnwidth]{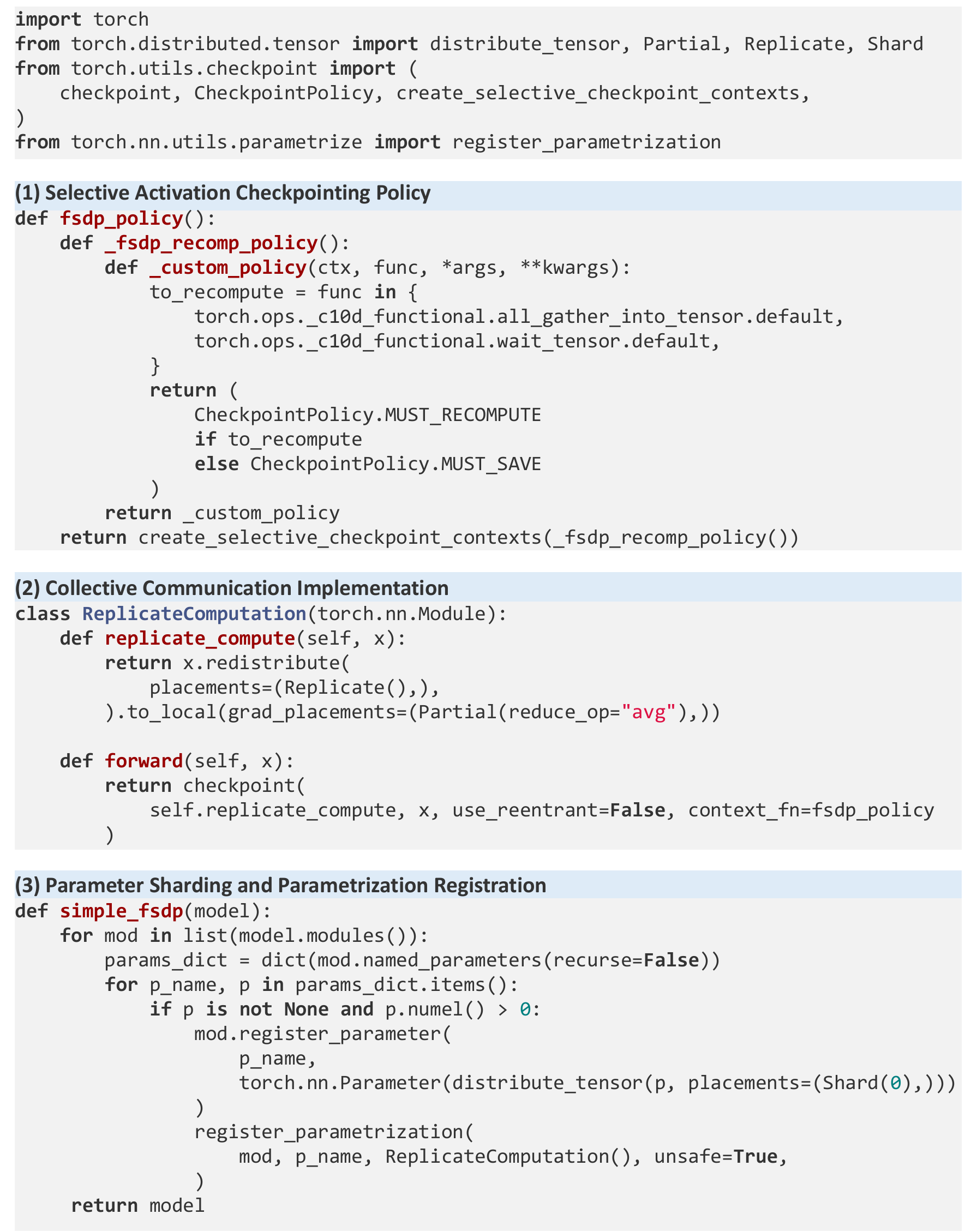} 
\end{center}
    \caption{\sys{}'s  frontend implementation.} \vspace{-0.3cm}
    \label{fig:allgather_recompute}
\end{figure}

In FSDP (or ZeRO-3), optimizer states, gradients, and model parameters are all sharded. The parameters are all-gathered in the forward pass and used for both forward and backward computation, whereas the computed gradients are reduce-scattered after backward computation for optimizer updates. \sys{} shards the parameters as DTensors during model initialization and utilize PyTorch primitive parametrization and DTensor API \texttt{redistribute} to implement the all-gather in the forward pass, as in Figure~\ref{fig:allgather_recompute}'s \texttt{ReplicateComputation}. Since DTensor's \texttt{redistribute} and parametrization are differentiable, the gradient reduce-scatter in the backward pass is automatically handled.

%In ZeRO-2, model parameters are sharded (in addition to the optimizer states in ZeRO-1), all-gathered in the forward pass for (both forward and backward) computation, and reduce-scattered after its backward usage for optimizer updates. In \sys{}, we shard the parameters as DTensors during model initialization, and utilize PyTorch primitive parametrization 
% and DTensor API \texttt{redistribute} to implement the all-gather in forward. Please see Figure~\ref{fig:allgather_recompute} for the code. In particular, since DTensor \texttt{redistribute} and parametrization are differentiable, the gradient reduce-scatter backward is automatically achieved.

As a memory optimization, in the forward pass, the all-gathered parameters used for computation can be immediately released afterwards to save memory; during backward, another all-gather is issued to regather the released parameters. Such semantics can be perfectly described using activation checkpointing, which releases activations after forward computation and recomputes them before being used in the backward pass. Hence, as in Figure~\ref{fig:allgather_recompute}'s \texttt{ReplicateComputation.replicate\_compute}, the parametrization takes the sharded parameters as input and replicates the parameters as activations, where selective activation checkpointing is employed to localize the checkpointing behavior of the FSDP-related communication operators.

The implementation benefits distributed training from two aspects: (i) \textbf{Simplicity}: users only need to wrap their model with \texttt{simple\_fsdp(model)}, and call \texttt{torch.compile} on the wrapped model. It allows the machine learning compiler to generate a full graph with both computation and communication operations for downstream optimizations; (2) \textbf{Debuggability}: the implementation does not alter the eager mode code execution. Users can still experiment and debug the model for agile development. 

% code : https://docs.google.com/document/d/1G1zrCmChihyYief6Zn0lZ8SpxHiJAnOKUq0EaVPMOKM/edit?tab=t.0

\begin{figure*}[!ht]
    \centering
    \includegraphics[width=\linewidth]{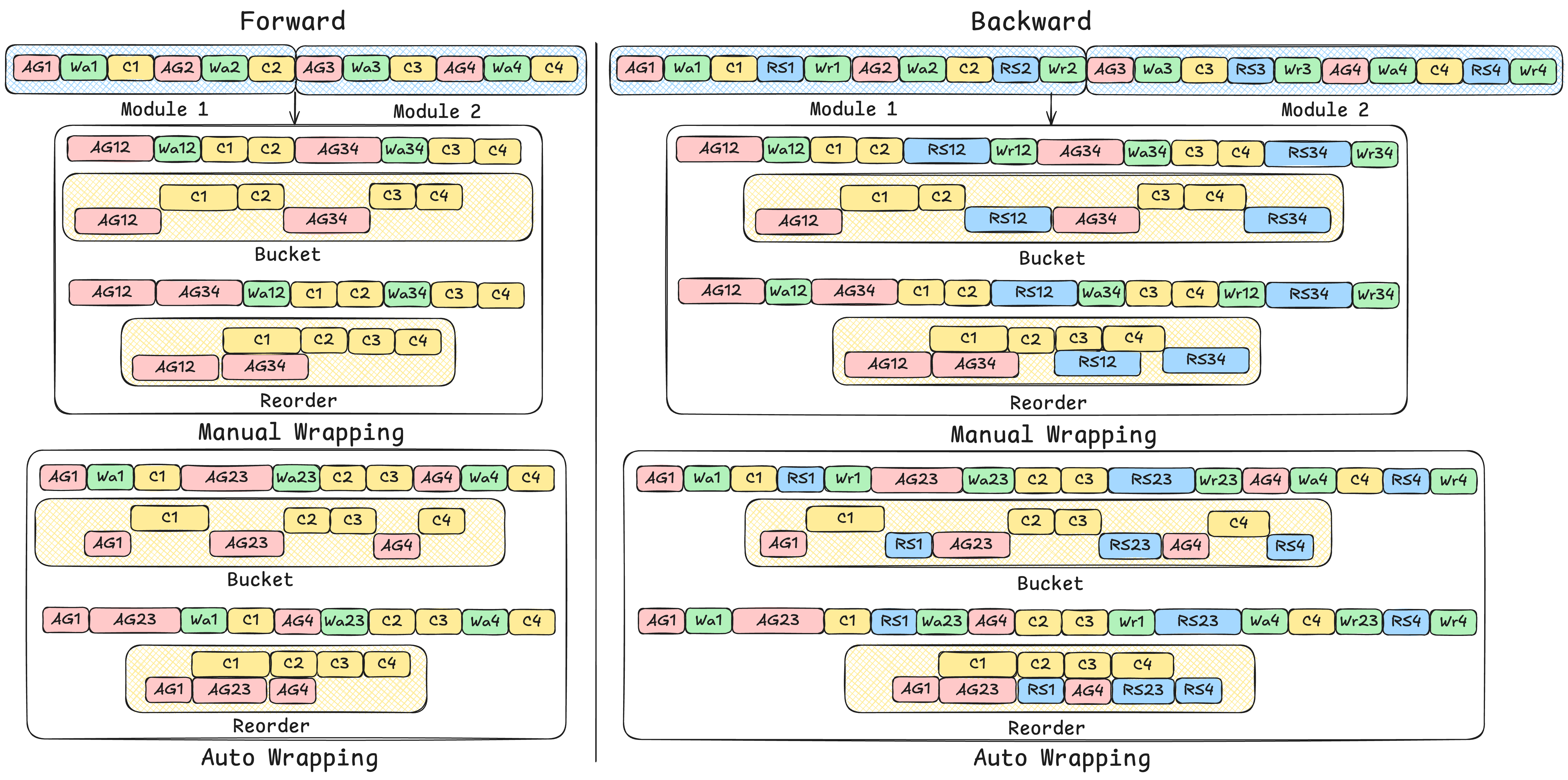}\vspace{-10pt}
    \caption{An overview of \sys's optimizations and model wrapping. The left side is the forward pass, and the right side is the backward pass. We show the IR node scheduling in \compiler{} and the corresponding execution order in GPU in the \textcolor{Dandelion}{yellow box}. The \textcolor{Cyan}{blue box} indicates the IR nodes are from the same module.
    In the \textbf{Manual Wrapping}, the all-gathers (AG) and reduce-scatters (RS) from the same module are bucketed as new communication IR nodes. Then, the bucketed communication and computation are reordered to enable communication prefetch during the current computation.  In the \textbf{Auto Wrapping}, the all-gather and reduce-scatter are bucketed as long as the bucketed communication can be overlapped by the current computation and does not exceed the memory limit. Then, the bucketed communication and computation are reordered to hide communication exposure. }\vspace{-10pt}
    \label{fig:overall}
\end{figure*}

\subsection{Optimizations}\label{subsec:optimization}
The graph traced from \sys{} is lowered to a set of IR nodes in \compiler{}. This alone does not yield optimized training performance, as the communication and computation operations from \texttt{ReplicateComputation} are shared per-parameter in sequential order, and all of the communication operations are exposed.
As depicted in Figure~\ref{fig:overall}, we introduce two optimizations in \compiler{} to enhance the distributed training performance: (1) \textbf{Bucketing} to group and merge communication IR nodes to reduce the frequency of issuing base communication; (2) \textbf{Reodering} to prefetch the communication IR nodes for overlapping with current computation.  

%After tracing the full graph, \compiler{} generates intermediate representation (IR) nodes for the model. For each child module wrapped with \texttt{RelicateCompute}, a corresponding all-gather and wait node is generated for the compute IR node. 

\subsubsection{Bucketing} \label{subsubsec:bucket}
The communication cost between two devices comprises a base latency for establishing the communication and a transmit latency proportional to the transmitted word size~\cite{nccl}. By bucketing the communication IR nodes, \sys{} issues the base communication once for all of the bucketed nodes and thereby reduces the overall communication time. 

As in Figure~\ref{fig:overall}, the individual all-gather/reduce-scatter IR node reads the data and issues the communication. To bucket the all-gather IR nodes $AG1$ and $AG2$, \sys{} allocate a bigger buffer that flattens and concatenates the tensor from each individual all-gather. The new all-gather $AG12$ and all-gather-wait $Wa12$ are created to gather the bigger buffers from other devices and copy out the gathered data based on their original tensor size. 

To bucket reduce-scatter, \sys{} splits the obtained gradient into chunks based on world size and concatenates the gradients from the individual reduce-scatter $RS1$ and $RS2$'s data to create a bigger buffer. A new reduce-scatter $RS12$ and reduce-scatter-wait $Wr12$ are created to average the buffer data gathered from other devices. The gradients are read out from $RS12$ to update the local model weight.

\subsubsection{Reordering} 
\label{subsubsec:reorder}
The collective communication all-gather and reduce-scatter are asynchronous, allowing it to occur concurrently with the computation on different CUDA streams. In Figure~\ref{fig:overall}, in the forward pass, each computation has an all-gather and an all-gather-wait to gather the data; in the backward pass, each computation has additional reduce-scatter and reduce-scatter-wait to update the gradient.
%The wait IR node blocks the process until the respective asynchronous communication finishes, ensuring the data is ready before use.
Reordering the IR nodes ensures the communications can overlap by the computations and thereby reduces the communication exposure. 
%To prefetch parameters for the next round's computation, we place the next round's collective communication before the current round's computation. As such, the communication is executed concurrently with the current computation on different CUDA streams. 

As in Figure~\ref{fig:overall}, we use the manual wrapping as an example; the reordering process is as follows: 
(1) In the forward pass, the $AG34$ is reordered in front of $Wa12$. It allows $AG34$ to overlap with compute $C1$;
(2) In the backward pass, the $AG34$ is placed after $Wa12$, enabling $AG34$ to overlap with $C1$. 
The $Wr12$ is placed before $RS34$, such that $RS12$ can overlap with the later compute  $C3$ and $C4$.

There are additional computations to copy out data from all-gather and reduce-scatter after the respective wait IR node. In the forward pass, by placing $AG34$ before $Wa12$, $AG34$ can further overlap with the compute to copy out data from the bigger buffer after $Wa12$. In the backward pass, $RS12$ is before $Wa34$, making it overlap with the compute to copy out data from $AG34$; thereby, $AG$ can be placed after $Wa$ as the copy-out compute has already been overlapped. 

%It saves the memory by placing $AG_{j+1}$ behind $Wa_{j}$, because the compute to copy-out $AG_{j+1}$ is already overlapped by $RS_j$.

% \begin{figure}
%     \centering
%     \includegraphics[width=0.95\linewidth]{mlsys2025style/figs/bucket.png}
%     \caption{Bucket all-gather/reduce-scatter nodes to reduce the inter-node communication frequency}
%     \label{fig:bucket}
% \end{figure}
%In the backward pass, the procedure is repeated for the reduce-scatter, where the copy-in buffer takes the $RS_{1-4}$'s data as input and creates a new reduce-scatter $RS_{new}$. The data is copied from the $RS_{new}$ to a copy-out buffer and update the model gradients.

\subsection{Model Wrapping}\label{subsubsec:wrapping}
Building on top of the optimizations in Section~\ref{subsec:optimization}, \sys{} provides two wrapping interfaces, namely, manual-wrapping and auto-wrapping, to bucket communication IR nodes together and reorder them for overlapping with computation operations.

The manual-wrapping buckets communication IR nodes based on pre-defined module lists. It provides the same functionality as those in FSDP2~\cite{fsdp-rfc}, where users can customize module wrapping after model definition. The auto-wrapping provides a more fine-grained and automatic bucketing interface, where no input from users is required. As the model is shared per parameter, \sys{} employs a greedy algorithm to atomically bucket communication IR nodes from each parameter for minimized exposure.

%IR nodes based on , which provides the same interface as 's eager mode. To experience the benefit of fine-grained per-parameter sharding, \sys{} provides auto-wrapping that automatically buckets communications at the parameter level and minimizes the exposed communication.  

%Building on top of reordering and bucketing, we introduce a module wrapping to wrap IR nodes from the same child module; and a auto-wrapping to predict IR node communication/computation times in \compiler{}. Then, it uses a greedy algorithm to bucket corresponding all-gather/reduce-scatter nodes and minimize the exposed communication. 

\subsubsection{Manual-wrapping}
In \compiler{}, each IR node contains metadata that traces its original module name. It enables \sys{} to construct a mapping between module names and their corresponding IR nodes.
Thus, users can customize the wrapping rules by providing a list of module names. \sys{} then wraps the communication/computation nodes between these modules. As in Figure~\ref{fig:overall}, the communication IR nodes from module 1 and module 2 are bucketed separately and reordered to overlap the bucketed communication.

%By default, \sys{} wraps each child module of the given model.  Users also have the option to define custom wrapping rules by 

%The IR nodes' metadata traces which child module the node comes from. By pattern-matching the IR node's source module and the user-defined wrapping plan, \sys{} buckets all-gather and reduce-scatters from the same group and issue a new all-gather and reduce-scatter for the group. Afterward, the bucketed IR nodes are reordered to overlap the newly created all-gathers and reduce-scatters.

\subsubsection{Auto-wrapping}

\textbf{Profiling}
The profiling algorithm estimates the IR nodes' communication and computation time in \compiler{}.
For the computation node, \sys{} converts the FakeTensor (containing tensor metadata without actual data) into real PyTorch Tensors. It executes the computation node's Python kernel with these real Tensors as input and records the CUDA event time $T_{c}$ and the peak memory $M_{c}$. 
For the communication node, we formulate the estimated communication time as $T_{m}=\alpha + \beta n$ where $n$ denotes the transmitted word size and $\alpha$, $\beta$ are the transmit parameters~\cite{nccl}. %The estimation is performed on rank0 and broadcast to synchronize the results across all ranks. 

\textbf{Wrapping} The wrapping algorithm automatically buckets the communication IR nodes to minimize communication exposure while keeping the memory within the limit. %as long as (1) the bucketed communication IR node can be overlapped by the current computation operations and (2) the computation of the pre-fetched parameters in the next step will not exceed the memory limit. 

\begin{table}[!ht]
    \centering
    \begin{tabular}{c|c}
    \toprule
       Variable  &  Definition \\\hline
       $T_{m}^{AG}$ & current step's bucketed AG's communication time\\
       $T_c$ & current step's computation time\\
       $M_c$ & next step's peak computation memory \\
       $T_{m}^{RS}$ & last step's bucketed RS's communication time \\
       $T_{mi}^{AG}$ & $i$-th AG's communication time  \\
       $T_{ci}$ & time to compute the parameters pre-fetched  by $i$-th AG  \\
       $M_{ci}$ & peak memory to compute the parameters pre-fetched  by $i$-th AG \\
       $T_{mi}^{RS}$ & time to reduce-scatter the gradient for parameters in $i$-th AG  \\
    \bottomrule
    \end{tabular}
    \caption{Variable definition}
    \label{tab:variable}
\end{table}

We show an example of the bucketing decision process in Algorithm~\ref{alg:auto_wrap}, where the variables are defined in Table~\ref{tab:variable}.
In the forward pass, we decide if the $i$-th all-gather node can be bucketed with the previous all-gather as long as it satisfies (1) Time Constraint: the communication time after bucketing the $i$-th all-gather, denoted as $T_{(m+mi)}^{AG}$, can be overlapped by the current step computation (line 4) and (2) Memory Constraint: the pre-fetched computation memory in the next step does not exceed the memory limit $M_{max}$ (line 5). Otherwise, the $i$-th all-gather will not be bucketed with the previous all-gather.

In the backward pass, the $i$-th all-gather nodes are bucketed with the previous all-gather as long as it satisfies (1) Time Constraint: the previous step's reduce-scatter $T_{m}^{RS}$ and the current step's all-gather when bucketing the $i$-th all-gather, denoted as $T_{(m+mi)}^{AG}$, can be overlapped by the current step's computation (line 10), and (2) Memory Constraint: the pre-fetched computation memory in the next step does not exceed the memory constraint $M_{max}$ (line 11). The corresponding reduce-scatter IR nodes of the all-gathers are bucketed as well.  
Otherwise, the $i$-th all-gather will not be bucketed with the previous all-gather.

\begin{algorithm}[!ht]
\caption{Auto Wrapping Algorithm}
\label{alg:auto_wrap}
\begin{algorithmic}[1]
\STATE  {\bfseries Input:} $T_{m}^{AG}$, $T_{m}^{RS}$, $T_c$, $M_{c}$, $T_{mi}^{AG}$, $T_{mi}^{RS}$, $T_{ci}$, $M_{ci}$
\STATE  {\bfseries Output: } \texttt{True} for bucket; \texttt{False} for not bucket
\IF{$isForward$}
\STATE $timeConstraint$ = ($T_{(m+mi)}^{AG}\leq T_{c}$)
\STATE $memConstraint$ = ($M_{c} + M_{c(i)} \leq M_{max}$)
\IF{$timeConstraint$ \AND $memConstraint$ }
 \STATE \textbf{return} \texttt{True}
\ENDIF
\ELSIF{$isBackward$}
\STATE $timeConstraint$ = ($T_{m}^{RS} + T_{(m+mi)}^{AG} \leq T_{c}$)
\STATE $memConstraint$ = ($M_{c} + M_{c(i)} \leq M_{max}$)
\IF{$timeConstraint$ \AND $memConstraint$ }
 \STATE \textbf{return} \texttt{True}
\ENDIF
\ENDIF
 \STATE  \textbf{return} \texttt{False}
\end{algorithmic}
\end{algorithm}

%\textcolor{red}{1. have a table for variables 2. have 2/3 instead of i 3. have a reference to which algorithm in line 4/7. }

\subsection{User interface} \sys{} provides simple plug-in-play interface. 
After defining the parallelism configs, users employ the \texttt{simple\_fsdp} API to wrap the model with \sys, as introduced in Section~\ref{subsec:overview}. Then, the \texttt{torch.compile} API compiles the model, tracing both communication and computation operations.  

The model wrapping, including reordering and bucketing, is handled by the \compiler{} backend. No additional modifications are required for the existing distributed training codebase. It greatly reduced the burden of maintaining distributed training code while providing performance gains.

\texttt{fullgraph=True} generates a full model graph. If the model has untraceable content, e.g., data-dependent control flow, setting \texttt{fullgraph=False} splits the graph into several subgraphs for \sys{} to optimize. 

\smallskip
\begin{lstlisting}[language=Python,numbers=left,style=pythonstyle]
torch._inductor.config.simplefsdp.bucket_mode = "auto"
torch._inductor.config.simplefsdp.enable_reorder = True
model = simple_fsdp(model)
model = torch.compile(model, fullgraph=True)
\end{lstlisting}

\section{Composability}\label{sec:composability}
\sys{} is natively implemented with \emph{DTensor}, \emph{parametrization} and \emph{selective activation checkpointing}, making it easy to be integrated with techniques in Section~\ref{subsec:dist_training} to train large models with few lines of code.

\textbf{Meta initialization} During model weight initialization on the meta device, \sys{} disables the all-gather parametrization to reduce the time and memory required to load the model.

\textbf{Mixed precision training} The \verb|param_dtype| and \verb|reduce_dtype| are parsed into the DTensor redistribution. During training, the model parameters are cast to \verb|param_dtype|, while the gradients are cast to \verb|reduce_dtype|. Mixed precision training is enabled by setting \verb|param_dtype| to the 16-bit floating point and \verb|reduce_dtype| to the 32-bit floating point.

\smallskip
\begin{lstlisting}[language=Python,numbers=left,style=pythonstyle]
def replicate_compute(self, x):
    output = x.redistribute(
        placements=(Replicate(),),
        forward_dtype=self.param_dtype,
        backward_dtype=self.reduce_dtype,
    ).to_local(grad_placements=(Partial(reduce_op="avg"),))
\end{lstlisting}

\textbf{Tensor Parallel} In parametrization computation, a model parameter can be initialized as a 2D DTensor, doubly sharded on both Data Parallel (DP) and Tensor Parallel (TP) dimensions. During computation, it is first redistributed (via an all-gather) on the DP sub-mesh, and then represented as a sharded DTensor on the TP sub-mesh, ready for the following TP computations.

% \begin{figure}[!ht]
%     \centering
%     \includegraphics[width=\linewidth]{mlsys2025style/figs/composability.pdf}
%     \caption{Compose \sys{} with mix-precision training and tensor parallel.}
%     \label{fig:composibility}
% \end{figure}

\textbf{Pipeline Parallel} The model is partitioned into several submodules, with each device receiving a copy of the assigned submodule. \sys{} wraps the received submodule before computation. No additional code is required to ensure compatibility with Pipeline Parallel.

\textbf{Activation checkpointing} Similarly, the activation checkpointing is applied before \sys{}. After defining which operations to recompute, \sys{} wraps the model and applies the additional checkpointing policies to the FSDP communication operations. No additional code is needed to compose with activation checkpointing.

% \begin{table}[ht]
%     \centering
%     \begin{tabular}{lcccc}
%     \toprule
%          &  MPT & AC & TP & PP\\\midrule
%      LoC &  1 & 2 & 8 & 0\\
%     \bottomrule
%     \end{tabular}
%     \caption{Caption}
%     \label{tab:my_label}
% \end{table}

% \paragraph{Model Wrapping.} The model wrapping is implemented in \compiler's scheduling phase after node fusion. \texttt{profile\_nodes} estimates the compute and communication node runtime and broadcasts the estimation from rank0 to other ranks. \texttt{auto\_bucket} uses these estimates to greedily bucket all-gather and reduce-scatter communication nodes and minimize communication exposure following Section~\ref{subsubsec:bucket}-~\ref{subsubsec:wrapping}. Then, the bucketed nodes are reordered to overlap with compute nodes following Section~\ref{subsubsec:reorder}.

% \smallskip
% \begin{lstlisting}[language=Python,numbers=left]
% # Bucketing
% run_time_dict = profile_nodes(self.nodes)
% self.nodes = auto_bucket(self.nodes, run_time_dict)
% # Reordering
% self.nodes = reorder_all_gather(self.nodes)
% self.nodes = reorder_reduce_scatter(self.nodes)
% \end{lstlisting}

\section{Experiments}
%In this section, we first introduce the experiment setups. Then, we demonstrate \sys's effectiveness, composability, and scalability in Section~\ref{subsec:performance}-\ref{subsec:auto}. Section~\ref{subsec:analysis} provides additional analysis regarding \sys's debuggability, convergence, compilation efficiency, etc. 

%\subsection{Experiment Setup}
\textbf{Infrastructure}  We build \sys{} in PyTorch \compiler{} with $\sim$ 2K LoC. The benchmarking is performed on TorchTitan~\cite{Liang2024TorchTitanOP}. Our evaluation environment includes a CPU/GPU cluster with 16 nodes. Each node has 8 NVIDIA H100 GPUs, and the intra-node is connected via NVLink~\cite{wei20239}.

\textbf{Target Models and Metrics} We evaluate \sys{} on Llama 3.1 series~\cite{dubey2024llama} models of various sizes. The details are in Table.~\ref{tab:model_details}. 

% config from : https://arxiv.org/pdf/2407.21783
\begin{table}[!ht]
    \centering
    \begin{tabular}{c|cccc}
    \toprule
       Model   & Layers & Model Dim. & FFN Dim. & Head Num. \\\midrule
      8B   & 32 & 4,096 & 14,336 & 32\\
     % 11B  & T+I & 32 & 4,096 & 14,336 & 32\\
      70B  & 80 & 8,192 & 28,672 & 64\\
      405B & 126 & 16,384 & 53,248 & 128\\
    \bottomrule
    \end{tabular}
    \caption{Llama 3.1 Model Configurations.}\vspace{-10pt}
    \label{tab:model_details}
\end{table}

The performance is evaluated by training the models on the C4 dataset~\cite{zhu2024multimodal}. The reported metrics are (i) \textbf{Token-per-second (TPS)}, the training throughput as the number of processed tokens per second; (2) \textbf{Memory}, the peak training memory. 

\textbf{Baselines} We compare \sys{} with PyTorch FSDP2~\cite{fsdp-rfc} developed by the official PyTorch team. It is an improved implementation of FSDP~\cite{zhao2023pytorch} by offering lower memory consumption and higher throughput.

$\bullet$ FSDP2-eager: The target model's submodules are wrapped as FSDP2 units and executed in the PyTorch eager mode. It is the widely adopted distributed training setting. 

$\bullet$ FSDP2-compile: Each target model's submodules is compiled with \compiler{} before being wrapped as an FSDP2 unit. It offers a stronger baseline by applying \texttt{torch.compile} to computation operations without handling collective communication tracing.

For fair comparisons, in Section~\ref{subsec:performance}-~\ref{subsec:compose}, \sys{} employs manual-wrapping to bucket the communication per transformer-block, and FSDP2-compile compiles the computation operations in each transformer-block region. 
In Section~\ref{subsec:auto}, we study the auto-wrapping performance.

\begin{figure*}[h]
    \begin{minipage}[t]{\linewidth}
    \vspace{0pt}
    \centering 
    \vspace{-0.3cm}{\includegraphics[width=0.5\textwidth]{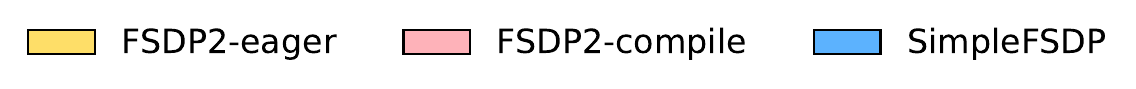}}\\
    \subfloat[8B (FSDP)]
    {\includegraphics[width=0.58\textwidth]{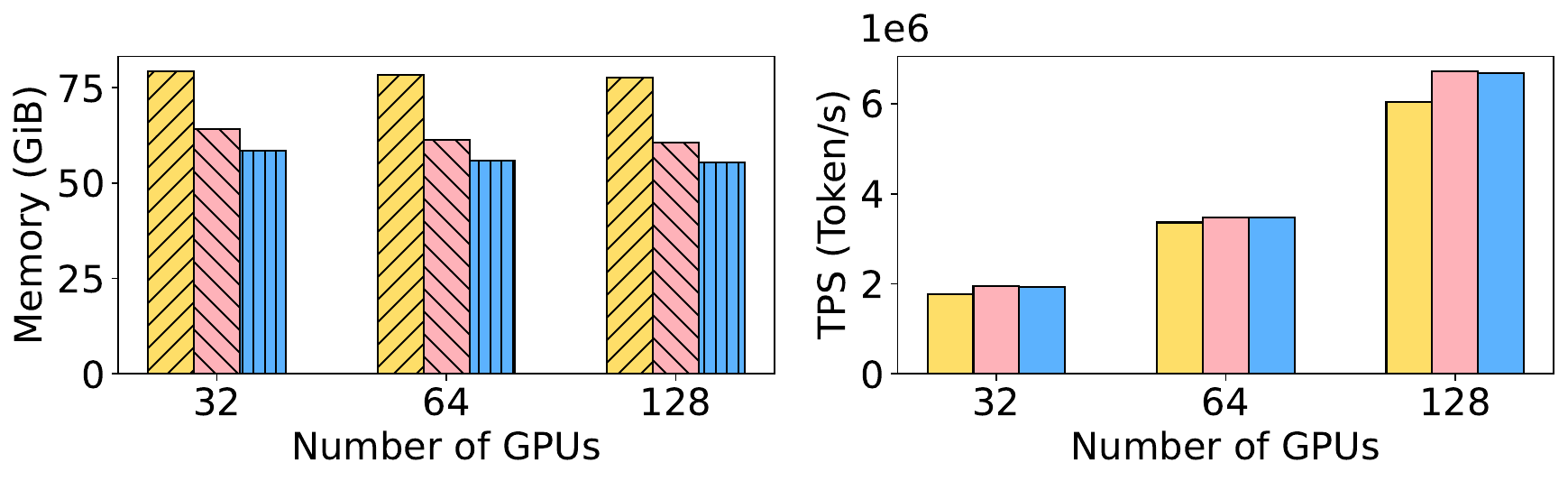}
    \vspace{-0.2cm}\label{subfig:8b1d}}
     \subfloat[70B (FSDP+TP+PP)]
    {\includegraphics[width=0.23\textwidth]{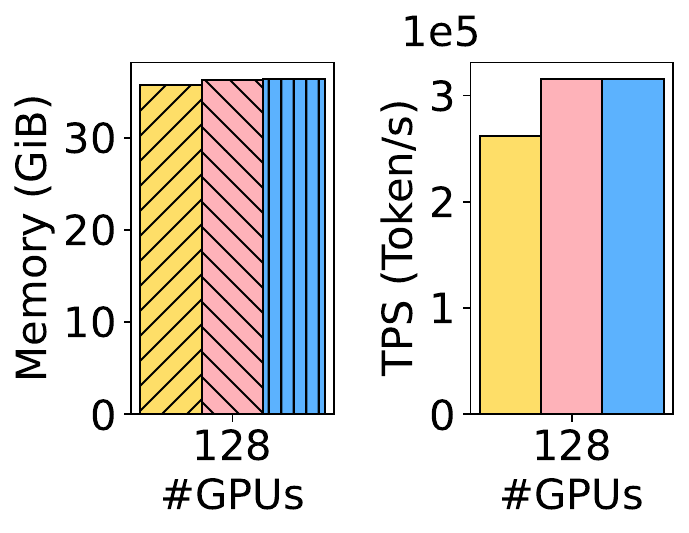}
    \vspace{-0.2cm}\label{subfig:70b3d}}
    \\
    \subfloat[8B (FSDP+TP)]
    {\includegraphics[width=0.58\textwidth]{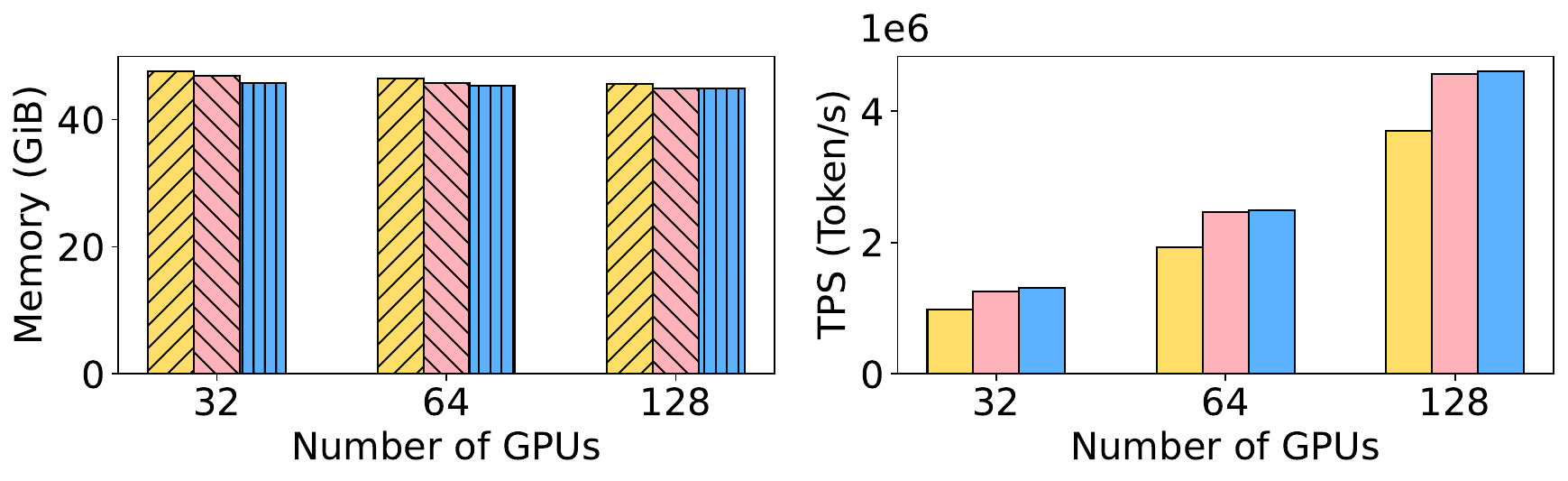}
    \vspace{-0.2cm}\label{subfig:8b2d}} 
\subfloat[405B (FSDP+TP)]{\includegraphics[width=0.23\textwidth]{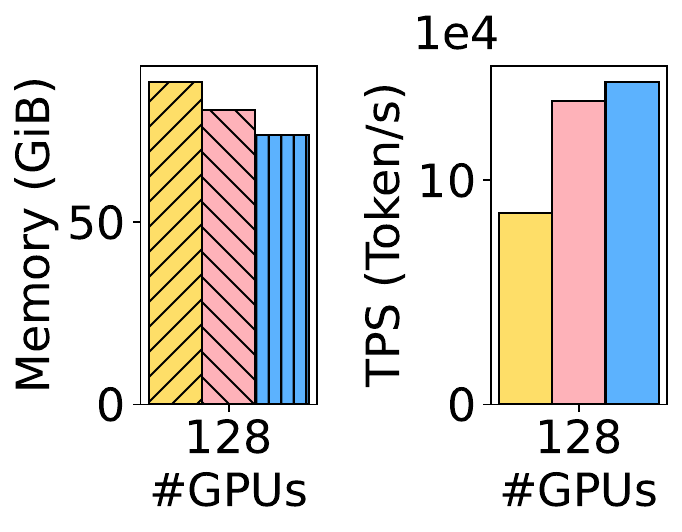}
    \vspace{-0.2cm}\label{subfig:405b2d}}
    \\
    \subfloat[70B (FSDP+TP)]
    {\includegraphics[width=0.58\textwidth]{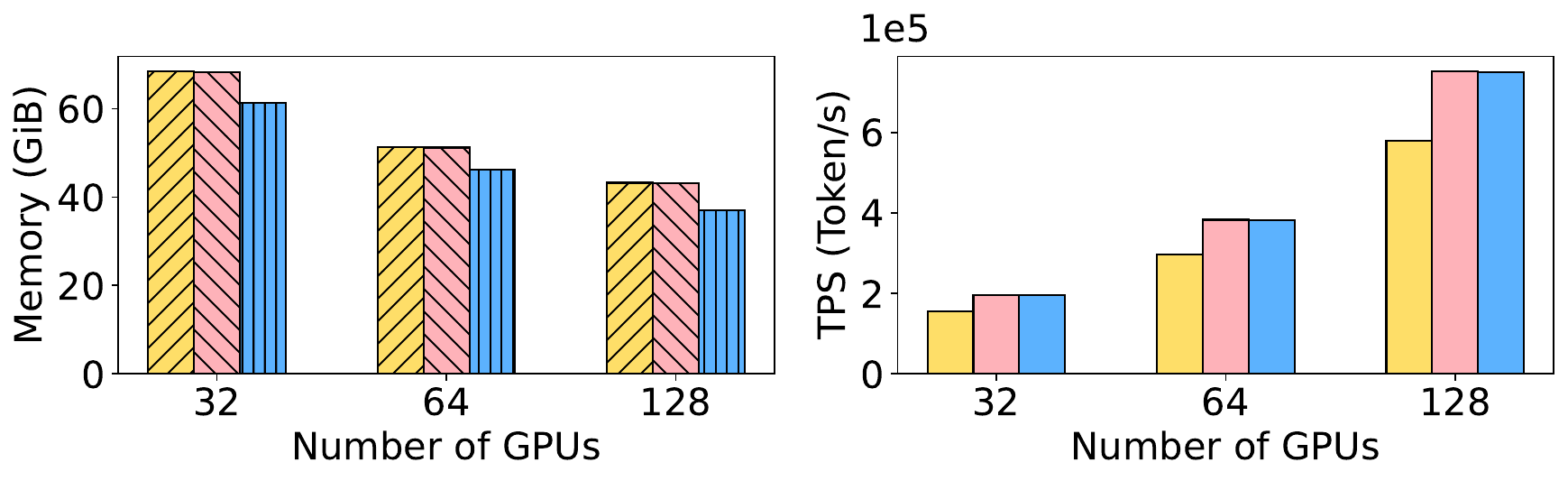}
    \vspace{-0.2cm}\label{subfig:70b2d}}
    \subfloat[405B (FSDP+TP+PP)]
     {\includegraphics[width=0.23\textwidth]{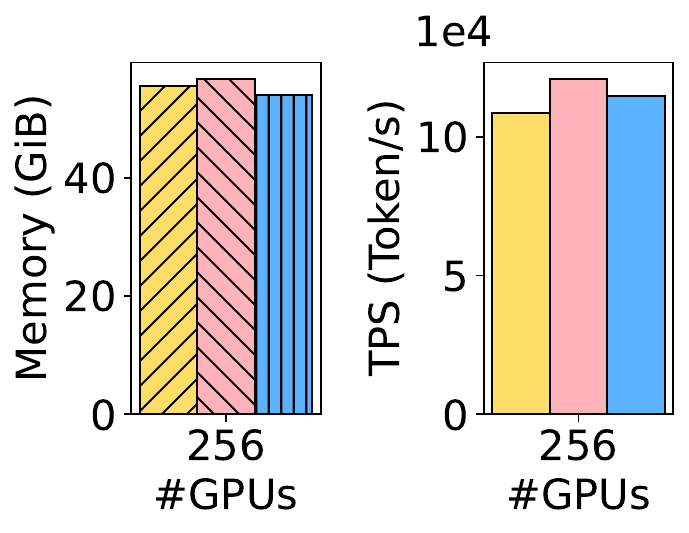}
     \vspace{-0.2cm}\label{subfig:405b3d}}
    \\
    \caption{\sys{} performance on LLaMA-3 8B, 70B, and 405B models when training on different numbers of H100 GPUs. We report the peak memory in GiB and the throughputs in TPS (tokens/s). }
    \end{minipage}\vspace{-10pt}
    \hfill
\end{figure*}
\subsection{\sys{} Performance}\label{subsec:performance}

%-\ref{subfig:13b1d} and Llama 3.2 11B trained with both FSDP and full activation checkpointing (AC) % 
Figure~\ref{subfig:8b1d} shows the memory and throughput for Llama 3.1 8B trained with FSDP on 32, 64, and 128 H100 GPUs.
Compared to FSDP2-eager, \sys{} on average saves 27.72\% peak memory and improves the throughput by 7.49\%. The performance gains are primarily from the memory optimizations, IR node fusions, etc, in \texttt{torch.compile} that make the training more efficient. Compared to FSDP-compile, \sys{} still maintains 8.65\% peak memory reduction by tracing the full-graph, where \texttt{torch.compile} obtains a global view and allocates the memory better. The transformer-based LLaMA models are computation-intensive, meaning the communication is fully overlapped when only FSDP is applied. As such, both \sys{} and FSDP2-compile hit the throughput upper bound for training the Llama 3.1 8B model, resulting in on-par throughput performance. %Note that, compared to FSDP2-compile, \sys{} traces the full model graph, which provides the optimization opportunities for 

% In this section, we demonstrate \sys{}'s effectiveness when training different models under different GPU configurations. As in Figure~\ref{}, when training the Llama 3.1 8B model using FSDP, \sys{} averagely saves 21.73GiB and 5.37GiB memory compared with FSDP2-eager and FSDP2-compile, respectively. 
% As Llama 3.1 8B is a computation-heavy model, the communications are fully overlapped in data-parallel settings. As such, both FSDP2-compile and \sys{} archive the computation upper-bound, where no communication is exposed during training. \sys{} and FSDP2-compile averagely archives $\sim$ 7.49\% throughput improvement over FSDP2-eager. The improvement primarily comes from the optimizations, e.g., IR node fusion in \texttt{toch.compile}.

%\textcolor{blue}{TODO: add llama-11B}

\textbf{Where memory savings come from} We identify the following three main reasons for \sys's memory savings.
(1) \sys{} works at a tensor-level, allowing a finer granular memory management (e.g. all-gathered parameters can be released sooner), compared with FSDP2’s module-level behavior. This gives memory advantage to \sys{} on the Llama model.
(2) \verb|torch.compile| makes different decisions on what activations to save for FSDP2 block-level compilation and \sys{} whole-model compilation. The FSDP2-compile will save additional transposed tensors from scaled dot product attention (SDPA) outputs, whereas \sys{} only saves SDPA outputs.
(3) \sys{} manages bucketing on the same CUDA stream as the rest computes. FSDP2 uses multiple streams, which has a throughput benefit but can cause memory allocation fragmentation.

\subsection{\sys{} Composability and Scalability} \label{subsec:compose}
In this section, we present the performance after composing \sys{} with Tensor Parallel and Pipeline Parallel.  All of the models employ full activation checkpointing (AC) and mixed precision training. In the compile mode, we apply Asynchronous Tensor Parallel~\cite{wang2022overlap}.

\textbf{2D Composability}  The Llama 3.1 8B and 70B models are trained with FSDP and Tensor Parallel~\cite{shoeybi2019megatron}.  The Tensor Parallel degree is set to 8, and the batch size is set to 16 and 8, respectively. 
Figure~\ref{subfig:8b2d} and \ref{subfig:70b2d} show the performance when training Llama 3.1 8B and 70B models on 32, 64, and 128 GPUs.

\sys{} can be integrated with Tensor Parallel without degrading the performance. 
As seen, when training the 8B model, compared to eager mode, \sys{} averagely saves 2.67\% peak memory and improves the throughputs by 29.35\%. Apart from the IR node fusion, \sys{} benefits from the Asynchronous Tensor Parallel~\cite{wang2022overlap} that overlaps the submatrix multiplication with communication operations.  
Compared to FSDP2-compile, by tracing the full-graph, \sys{} further demonstrates 2.09\% throughput improvement. 

As the model size becomes larger, the full graph traced by \sys{} provides more memory optimization opportunities in \texttt{torch.compile} and yields better performance. 
As in the 70B model, \sys{} improves FSDP2-eager's throughputs by 28.26\% and reduces the training peak memory by 11.61\%. While both \sys{} and FSDP-compile hit the throughput upper bound, \sys{} further reduces 11.40\% peak memory from the full-graph tracing. 

%averagely saves 1.25GiB and 0.56GiB memory compared with FSDP2-eager and FSDP2-compile on Llama 3.1 8B model and 6.15GiB and 6.02GiB memory on Llama 3.1 70B model. 

\textbf{3D Composability} The Llama 3.1 70B models are trained with  FSDP, Tensor Parallel~\cite{shoeybi2019megatron}, and Pipeline Parallel~\cite{huang2019gpipe}. The Tensor Parallel degree is set to 8, the Pipeline Parallel degree is set to 8, and the batch size is set to 16.

\sys{} can be integrated with Pipeline Parallel without performance degradations. As seen in Figure~\ref{subfig:70b3d}, \sys{} improves the throughput by 20.31\% compared to FSDP2-eager while incurring less than 1GiB peak memory overhead. The models are partitioned into smaller submodules, with each device receiving a subgraph of the full model. \sys{} optimizes a smaller graph compared to 2D settings, thus achieving comparable performance with FSDP2-compile. Notably, \sys{} traces a communication-computation partitioned subgraph on each device, making it possible to overlap the bubbles from 3D training and bring opportunities for future optimizations. 

% While achieving comparable throughputs, \sys{}  reduces 7.74\% peak memory compared to FSDP2-compile. 

\textbf{Scalability} We show the Llama 3.1 405B model 2D parallelism training performance in Figure~\ref{subfig:405b2d} and 3D parallelism performance in Figure~\ref{subfig:405b3d}. The Tensor Parallel degree is set to 8, the Pipeline Parallel degree is set to 16, and the batch size is set to 2 in 2D parallelism and 16 in 3D parallelism.

\sys{} is scalable and maintains the performance enhancement when training ultra-large models. 
As seen, compared to FSDP2-eager, \sys{} improves the throughput by 68.67\% and 5.66\% on 2D and 3D parallel, respectively. Besides, \sys{} reduces the memory by 16.26\% and 2.64\% on 2D and 3D parallelism. 
Compared to FSDP2-compile, by tracing the full model graph, \sys{} improves the throughput by 6.06\% on 2D parallel and reduces the memory by 8.37\% and 4.63\% on 2D and 3D parallel, respectively.

The throughput gains and memory savings become more significant when training large models at scale: (1) Large model training requires millions of GPU hours~\cite{dubey2024llama,chowdhery2023palm}, thereby even small per-iteration throughput gains substantially reduce overall training time.; (2) The memory savings allow for larger batch sizes training per iteration, which in turn increases the throughput.

\subsection{Auto-Wrapping Performance}\label{subsec:auto}
\begin{figure}[!ht]
    \centering
    
    \vspace{-0.3cm}
{\includegraphics[width=0.5\columnwidth]{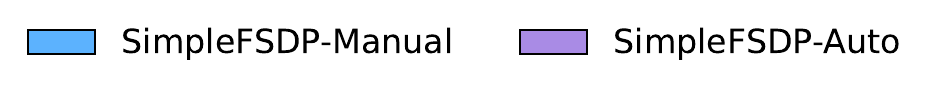}}\\
    
    \subfloat[Llama 3.1 8B]{
    \includegraphics[width=0.3\columnwidth]{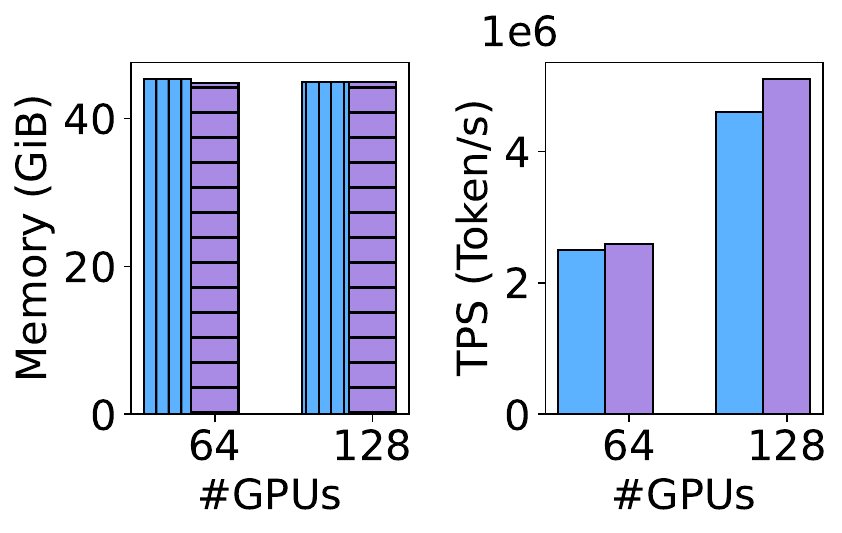}}
    \subfloat[Llama 3.1 70B]{
    \includegraphics[width=0.3\columnwidth]{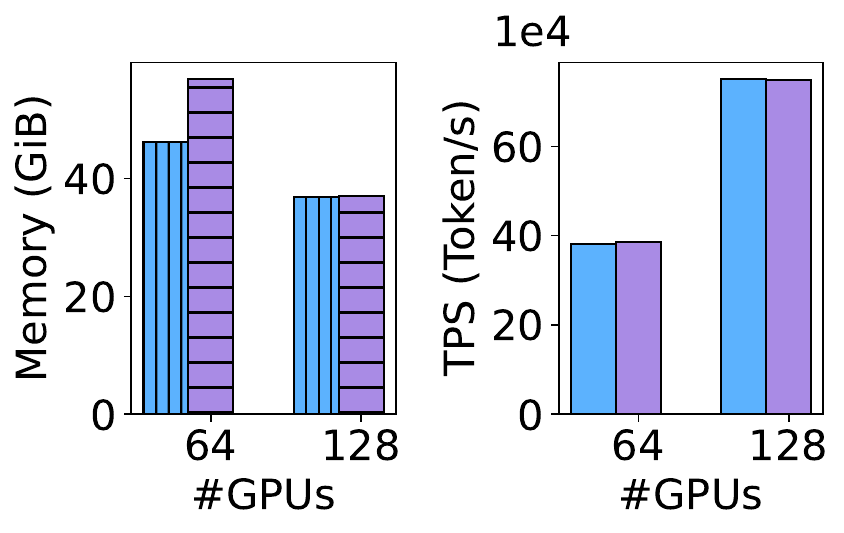}}
    \caption{Auto-Wrapping performance when training Llama 3.1 8B and 70B models on different numbers of H100 GPUs.} \vspace{-0.3cm}
    \label{fig:auto}
\end{figure}

The auto-wrapping performance is in Figure~\ref{fig:auto}. The communication operations are fully-overlapped when training Llama models with only FSDP. Hence, We show the performance when training Llama 3.1 8B and 70B on 2D parallelism, where Asynchronous TP communication is exposed. Other settings follow Section~\ref{subsec:compose}.

\sys-Auto reduces the exposed communication during large model training and does not require manual wrapping plans defined by the users. 
As seen, in 8B model, \sys-Auto achieves $\sim$ 7.34\% throughput improvement over \sys-Manual while maintaining comparable memory consumption. It means more communications are overlapped by \sys-Auto, providing both automation and performance enhancement to users. 

However, we also provide one case where \sys-Auto provides 0.8\% throughput improvement when training Llama 3.1 70B models on 64GPUs but incurs 10.61GiB memory overhead. It is primarily because \sys-Auto prioritizes minimizing the exposed communication, and the memory threshold we set is larger than the peak memory in \sys-Manual. As a result, \sys-Auto gives a suboptimal solution and scarifies the memory for throughput improvement. 
The major focus of \sys{} is providing an elegant way of tracing a full graph with both communication and computation operations for downstream applications. We leave exploring algorithms to generate more optimal overlapping plans as future work. 

\subsection{Analysis and Ablation Study}\label{subsec:analysis}
This subsection presents analysis of how different optimization components impact \sys's performance. By default, we train the Llama 3.1 8B model on 8 H100 with only FSDP and set the batch size to 1. Additional ablation studies are in the appendix.

\textbf{Debuggability} Apart from the compile mode performance gains, \sys{} exhibits usability in the PyTorch eager mode. It offers users the flexibility to print variables and experiment with various building blocks for debugging and agile development. 
As is shown in Table~\ref{tab:debug}, \sys{} achieves comparable memory consumption and throughput to FSDP2, which is primarily developed for the eager mode\footnote{\sys{} with no AC incurs slightly higher memory than FSDP2 due to the frontend design choice to make the codebase simpler and is not fundamental; after all, we would rely on the compiler for good performance.}. Notably, \sys{} offers eager-mode debuggability with greater simplicity, composability, and performance gains in compile mode.

\begin{table}[!ht]
    \centering
    \begin{tabular}{l|c|cc}
     \toprule
     Method & AC & TPS $\uparrow$ (Token/s) & Memory$\downarrow$ (GiB)\\\midrule
     FSDP2-eager& None & 47,088 & 86.75 \\ 
     \sys & None &46,936 & 91.91\\
     FSDP2-eager& Full & 37,504 & 37.35 \\ 
     \sys & Full & 38,504 & 29.80\\
    \bottomrule
    \end{tabular}
    \caption{\sys's debuggability in the eager mode. AC is for activation checkpointing.}
    \label{tab:debug}
\end{table}

\begin{figure}[!ht]

    \centering
    \includegraphics[width=0.6\linewidth]{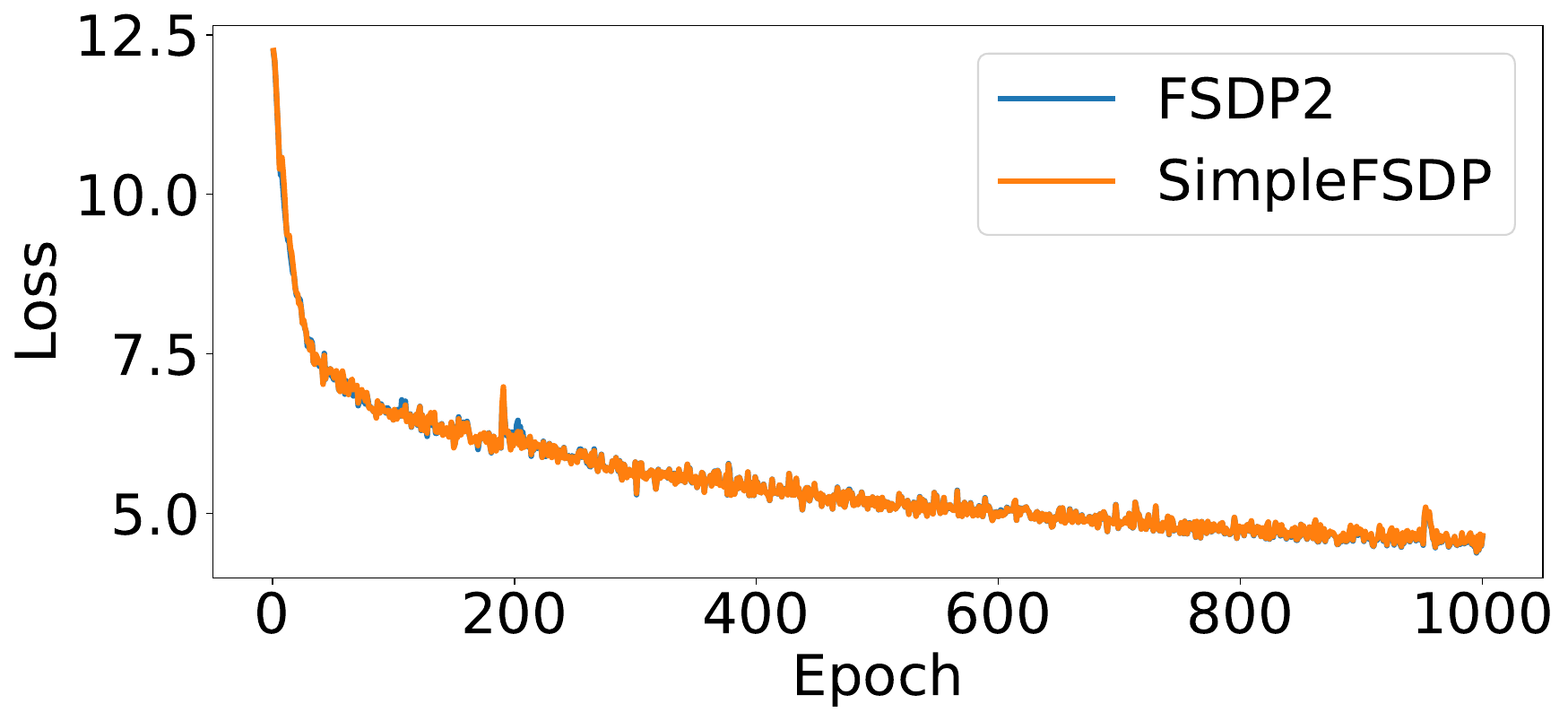}
    \vspace{-0.3cm}
    \caption{Loss curve of FSDP2 and \sys{} on Llama 3.1 8B.}
    \label{fig:loss}
\end{figure}

\textbf{Training convergence} \sys{} and its optimization components will not alter the training convergence. Figure~\ref{fig:loss} compares the loss plots of FSDP2 and \sys{} when training the Llama 3.1 8B model for 1,000 epochs on 8 H100 GPUs. As seen, the similar loss convergence for both methods demonstrates that \sys{} maintains model convergence and training stability.

\textbf{Compilation time}  The time takes to compile the Llama 3.1 8B model is in Table~\ref{tab:compile}. We split the total compilation time to reorder and bucket IR nodes in \sys{}, and the rest time to compile the model in \compiler{}. 
As seen, compiling \sys{} incurs negligible overhead compared to the overall training time, making it efficient.

\begin{table}[!ht]
    \centering
    \begin{tabular}{l|cccc}
    \toprule
         & Bucket & Reorder & Others & Total \\\midrule
     % FSDP2   &  - & - & \\
      \sys-Manual & 0.71 & 0.13 & 23.64 & 24.48 \\
      \sys-Auto & 3.87 & 0.16 & 21.53 & 25.56\\
    \bottomrule
    \end{tabular}
    \caption{Compilation time (in second) on Llama 3.1 8B.}
    \label{tab:compile}
\end{table}

\textbf{The effectiveness of reorder and bucket} Table~\ref{tab:reorder_bucket_ablation} shows the impact of reordering and bucketing on single and multi-node training. In single-node, reordering enables the computation and communication to happen concurrently. It increases the training throughput and memory usage. Bucketing further increases memory by grouping the IR nodes for computation and communication. However, it slightly reduces throughput due to the additional time needed for copy-in/copy-out data from the buffer, which is more significant compared to the intra-node base latency that bucketing aims to optimize.

In the multi-node setting, reordering similarly increases the throughput and memory compared with the vanilla setting. Bucketing further increases the throughput by merging the IR nodes to reduce the frequency needed to establish inter-node base communication, which is non-negligible compared to intra-node base latency.

\begin{table}[!ht]
    \centering
    \begin{tabular}{l|cc|cc}
    \toprule
        & \multicolumn{2}{c|}{1 node}  & \multicolumn{2}{c}{8 nodes} \\\cline{2-5}
         &  \begin{tabular}[c]{@{}c@{}}TPS $\uparrow$ \\ (Token/s)\end{tabular}& \begin{tabular}[c]{@{}c@{}}Memory$\downarrow$ \\  (GiB)\end{tabular}&  \begin{tabular}[c]{@{}c@{}}TPS $\uparrow$ \\ (Token/s)\end{tabular} & \begin{tabular}[c]{@{}c@{}}Memory$\downarrow$ \\ (GiB)\end{tabular}\\\hline
        vanilla & 50,976 & \textbf{67.26} & 333,440 & \textbf{56.42}\\
     + reorder  & \textbf{54,544} & 68.72 & 404,032 & 57.88\\
     + bucket &  49,168 & 69.06 & 405,632 & 58.15\\
     + reorder \& bucket & 52,480 & 69.08 & \textbf{428,352} & 65.74\\
    \bottomrule
    \end{tabular}
    \caption{Effectiveness of reorder and bucket.}
    \label{tab:reorder_bucket_ablation}
\end{table}

 \paragraph{The effectiveness of reordering all-gather before/after the last all-gather-wait} We analyze the impact of reordering all-gather before or after the last all-gather-wait in Table~\ref{tab:ag_reorder_ablation}.  In the forward pass, placing all-gather before the last all-gather-wait results in higher throughputs, as the compute to copy-out data from the last all-gather is overlapped by the reordered all-gather.

 In the backward pass, placing reduce-scatter-wait before the next reduce-scatter already optimizes the throughput gains that could have been achieved by the all-gather. Therefore, placing the all-gather after the last all-gather-wait will save the memory slightly.

\begin{table}[!ht]
    \centering
    \begin{tabular}{cc|cc|cc}
    \toprule
\multicolumn{2}{c|}{Forward}  & \multicolumn{2}{c|}{Backward}   & \multirow{2}{*}{\begin{tabular}[c]{@{}c@{}}TPS $\uparrow$ \\ (Token/s)\end{tabular}} & \multirow{2}{*}{\begin{tabular}[c]{@{}c@{}}Memory$\downarrow$ \\  (GiB)\end{tabular}} \\\cline{1-4}
before & after & before & after & &    \\\hline
\cmark & & \cmark &  & 51,912 & 69.08\\
\cmark & &  & \cmark & \textbf{52,480} & \textbf{69.08}\\
& \cmark & \cmark &  & 51,680 & 68.09\\
& \cmark & & \cmark & 51,776 & 68.09\\
    \bottomrule
    \end{tabular}
    \caption{The effectiveness of reordering all-gather before/after the last all-gather-wait.}
    \label{tab:ag_reorder_ablation}
\end{table}

\section{Discussion}
In this section, we discuss \sys{} in different use cases and potential future work.

\textbf{Graph breaks in model tracing}  While \sys{} obtains a full graph with both communication and computation operations, it does not require users to write code adhering to strict compilation constraints (e.g., avoiding data-dependent control flow or variable printing). 
Built upon \texttt{torch.compile}, when encountering non-traceable operations, the full graph is split into several subgraphs, each with communication and computation operations. \sys{} then optimizes each subgraph individually.

\textbf{Limitation} While \sys{} demonstrates promising performance, there are some cases where the auto-wrapping yields slightly worse performance than manual-wrapping. We found the discrepancy is due to inaccurate communication time estimation. Currently, the profiling algorithm only models the transmitted word size, whereas other factors like network topology~\cite{nedic2018network} are not taken into consideration. 
Besides, the greedy algorithm does not consider the overall nodes' runtime, which might result in suboptimal solutions. We leave these as future work to further improve \sys's auto-wrapping algorithm. We note that such accurate runtime estimations are not \sys-specific but would benefit any estimation-based algorithmic decision-making, e.g., auto-parallelism.

\textbf{Future work} Tracing a full graph with computation and communication operations brings the potential for many downstream works. 
For instance, researchers can reduce the exposed communication bubbles~\cite{he2021pipetransformer,zhang2023adagl,feng2024optimus} in distributed training or heterogeneous environments by auto-wrapping the computation/communication operations. 

\sys{} optimizes training performance by bucketing and reordering the IR nodes to minimize the communication exposure in multi-dimensional parallelisms. 
Given that our work sets up the necessary infrastructure to enable such communication optimizations in PyTorch compiler, we hope to promote more research in the related area. 
For example, with the type of computation-communication overlapping for Data Parallel in \sys{}, auto-parallelism engines~\cite{zheng2022alpa,chen2024slapo,lin2024nnscaler} can now aim for better plan execution and thus improved decision-making.

%By decoupling the tensor sharding with the parameter pre-fetching, \sys{} provides the opportunities to incorporate FSDP as one dimension within the auto-parallelism~\cite{zheng2022alpa,lin2024nnscaler} search spaces for auto sharding. The benefits of fine-grained communication and computation overlapping from pre-fetching can still be experienced by handling them in the \compiler{} backend. 

%While this work demonstrates \sys{}'s effectiveness in static settings, tracing a full graph with both communication and computation nodes can benefit auto-parallelism~\cite{zheng2022alpa,chen2024slapo,lin2024nnscaler}.

%It enables researchers to obtain more accurate and fine-grained node runtime estimations for optimizing the parallelism configs. %Besides, users do not need to manually insert the sharding components between the compute nodes, which further improves simplicity. 

\section{Conclusion}
We present \sys{}, a PyTorch-native compiler-based FSDP framework. It features simplicity for distributed training codebase maintenance, composability with other efficient training techniques, performance enhancement from full graph tracing, as well as debuggability and programmability from the PyTorch eager mode.
Building on top of the unique parametrizations implementation of all-gather to checkpoint parameters, \sys{} buckets and reorders the IR nodes for minimized communication exposure and provides customized and automated model wrapping interfaces to users.  
Extensive evaluations demonstrate \sys's efficacy in throughput gains, memory saving, and scalability toward tracing ultra-large models. 

\section*{Acknowledgements}

We sincerely thank Jason Ansel, Gregory Chanan, Soumith Chintala, Will Constable, Patrick Labatut, Gokul Nadathur, Damien Sereni, and Peng Wu for their support in making this work happen.

\clearpage
\newpage
\bibliographystyle{assets/plainnat}
\bibliography{paper,example_paper}

\end{document}